# AGOCS – Accurate Google Cloud Simulator Framework

Leszek Sliwko
Distributed and Intelligent Systems Research Group
University of Westminster
London, United Kingdom
Leszek.Sliwko@my.westminster.ac.uk

Vladimir Getov
Distributed and Intelligent Systems Research Group
University of Westminster
London, United Kingdom
V.S.Getov@westminster.ac.uk

*Abstract*—This paper presents the Accurate Google Cloud Simulator (AGOCS) – a novel high-fidelity Cloud workload simulator based on parsing real workload traces, which can be conveniently used on a desktop machine for day-to-day research. Our simulation is based on real-world workload traces from a Google Cluster with 12.5K nodes, over a period of a calendar month. The framework is able to reveal very precise and detailed parameters of the executed jobs, tasks and nodes as well as to provide actual resource usage statistics. The system has been implemented in Scala language with focus on parallel execution and an easy-to-extend design concept. The paper presents the detailed structural framework for AGOCS and discusses our main design decisions, whilst also suggesting alternative and possibly performance enhancing future approaches. The framework is available via the Open Source GitHub repository.

## I. Introduction

Correctly characterizing user behavior is of utmost importance when modeling Cloud workloads [14, 19]. Cloud workloads have been researched thoughtfully and are relatively well understood [15, 16, 24, 27]; however there have been limited attempts to accurately simulate Cloud workloads with consideration of detailed task parameters and constraints [6, 7, 10, 13], especially with consideration of workload scheduling [25].

Evaluating the performance of distributed applications and services without unrestricted access to existing Cloud environments is a very difficult task, which can also be addressed via simulation. Existing Cloud simulators do succeed in representing high-view infrastructure parameters (i.e. nodes, tasks and their dependencies), but they do not provide realistic and fine-grained system traces such as application and canonical memory, local and remote disk space, disk I/O, cycles per instruction, etc. This problem is even more acute when we consider deep system-critical mechanisms like tasks scheduling or fault handling and prevention schemes. Normally, Cloud providers would not allow developers to alter core system components such as the scheduler or the provisioning services in a working system.

The ideal scenario would allow each Cloud system designer to have unconstrained access to a considerable-sized Cloud system, which could be used as a test bed for developing models and strategies. However, the real case is that a developer has to compete for access to a computing center with many other business units.

Therefore, in our research, we have focused on building a flexible Cloud workload simulation framework, which could be deployed in a local environment (i.e. the researcher's desktop machine or laptop), while providing at the same time high-quality, detailed and accurate workload parameters of the simulated Cloud system. Previous research and analysis of available workload traces show that usually Cloud workloads are highly variable and non-cyclical. Spread around the globe, Cloud users are not constrained by predefined schedules, thus, in contrast to Grid and Cluster environments, the workload is not correlated to season or time of day [16]. Therefore, in order to research the proposed problem, a realistic Cloud workload simulation model is required. In order to setup a realistic scenario, two approaches can be used:

- Use an artificial Cloud workload generator as described in [6, 10, 24].
- Acquire and parse real-world workload traces [5, 8, 9, 12, 26] to a format, which can be used in further research.

A number of simulators already exist addressing various aspects of the current Cloud systems such as computations (CloudAnalyst [25], CloudSim [11]), networking/energy use (GreenCloud/NS2 [13] platform). However, those simulators, while very flexible, do not provide details about fine-grained parameters (i.e. memory page size, cache size, disk I/O time, cycles and memory access per instruction, etc.) that might be required in some types of simulation. Additionally, those simulation frameworks do not provide realistic resource usage statistics, which may actually be much lower than requested resource levels. Users tend to overestimate the amount of resources that they require, wasting in some cases up to 98% of the requested resource [16]. Such low-level parameters can be obtained only from real workload traces.

According to previous research and our own experience, building a high-fidelity workload generator is an extremely difficult task. A number of dependencies, constraints and other details to capture the overall dynamicity of Cloud systems [16, 27] forces researches to simplify their models and make assumptions. Therefore, we have decided to base our simulation approach and framework on real-world workload. The earlier high-fidelity simulators based on Google Cluster traces were built before the Omega work [20]. However, they were highly specialized for a certain type of research such as scheduling algorithms and are not publicly available. By contrast, our novel high-fidelity Cloud workload AGOCS simulator can be extended for a variety of research purposes and crucially it is readily available as an OpenSource project. We are not aware of any other similar tool providing comparable flexibility and fine-grained traces as found with a Cloud workload.

The remainder of this paper is organized as follows. In Section II we briefly describe several workload traces archives that could be used as a base for simulation purposes. In Section III we introduce our simulation framework design in detail and highlight the main design decisions. Section IV describes a typical use case scenario as used in our project. In Section V we suggest alternative designs that might improve simulator performance. Section VI compares the presented framework to existing solutions. Section VII evaluates the performance of AGOCS against its closest alternative – the CloudSim package [11]. Section VIII summarizes our experiments and presents our conclusions.

The created framework is available as Open Source project in the GitHub repository[1] and it is still under development. At the time of writing, the latest committed version was from 28th April 2016.

## II. WORKLOAD TRACES ARCHIVES

To the best of our knowledge, the following workload traces are publicly available at present:

- Google Cluster Data (GCD) project [14] – this repository includes detailed traces over a month-long period (May 2011) from a 12.5K-node network. The statistics include CPU usage, memory usage, disk I/O operations (for first two weeks, after that the logs configuration changed), network speed, etc.
- Grid Workload Archive [5] – hosted in Delft University of Technology in the Netherlands. This repository contains workload traces from almost a dozen grid systems. Most of them include CPU usage, memory usage, and disk I/O operations.
- Parallel Workloads Archive [9] – this repository contains over 30 workload logs from around the world. The earliest traces are from 1993, the latest are from 2012 (Intel NetBatch pools A-D grids). Those workload traces were thoughtfully cleaned of anomalies and data errors.
- MetaCentrum Workload Log [8] – archive contains data sets generated from TORQUE workload traces, deployed in the Czech National Grid Infrastructure MetaCentrum (22 clusters having 219 nodes with 1494 CPUs).
- Yahoo! M45 Supercomputing Project [26] – Yahoo! made its 4000-node Hadoop cluster's workload traces freely available to selected universities for academic research.

All of the above repositories with the exception of Yahoo! workload logs could be used freely for research as there are no legal restrictions and/or requirements for presenting this data or derived data in any kind of research work.

For the purpose of this research, the real-world workload traces from the GCD project are used. The main reason for the selection of this repository is the high quality of workload traces. Traces are complete and contain a low number of anomalies, which are thoughtfully explained including their schema and format [18]. They have been gathered from a large system over a significant period of time. Google offers a variety of services; therefore their backend systems are diversified and represent a complete spectrum of computation requirements. Finally, Google Inc. is a global company and its data centers are working continuously 24 hours a day – its clusters workloads are not cyclical, which could be a problem if we were using traces from a more local data center. The GCD workload traces are stored in the Cloud Storage Service [2] in the bucket 'clusterdata-2011-2'. The gsutil tool can download the traces; the compressed archives are approximately 41 GB, while the uncompressed archives are about 191 GB. Unfortunately, no logging system is perfect and every workload trace we examined contained a portion of anomalies. The GCD logs are of high quality; however there are a few known inconsistences [4]:

- Disk time data is not logged after the first 14 days due to changes in the monitoring system.
- Approximately 0.003% of jobs are not listed as they run on nodes not included in the workload traces.
- Approximately 70 jobs have no task information. The explanation given is that those jobs run but the tasks were disabled.
- Approximately 0.013% of the task events and 0.0008% of the job events have missing fields.
- Some resource statistics data are inaccurate – i.e. the cycles per instruction and the memory accesses per instruction parameters have values out of range for the underlying micro-architecture. Those parameters might be also affected by other tasks running on the same machine; however we believe this to be acceptable for simulation purposes.

Additionally, GCD traces were obfuscated from user data, operating system and platform details, job purpose information and special constraints names and values. These characteristics would be a very interesting point of research. It is also important to point out that GCD workloads are delayed by 10 minutes (60,000,000 micro seconds). This shift has been applied in order to split pre-existing cluster conditions such as already existing nodes from new incoming requests (i.e. scheduled tasks, resource utilizations, etc.). For further reading and a detailed analysis of GCD traces see [15, 16, 18, 19, 27].

## III. SIMULATION FRAMEWORK DESIGN

The major requirement of designing any distributed system simulator is accuracy. Interaction between networked entities can be very complex and it's of upmost importance to model them and well as timing of those changes correctly. In AGOCS, primary source of changes in system state are pre-recorded events. Available GDC data spans over a period of a month and all events records come with accurate timestamp information.

---

[1] https://github.com/lsliwko/masb

There exist several sources of events in GDC workload traces. All workload traces come in similar formats, where a change in the environment state is reported as an event. In GCD, the jobs queue is the base of all processing. All entries in the jobs queue traces relate to jobs submissions, jobs cancellations, changes in jobs' priorities, etc. Listed jobs contain a series of tasks (reported in separate log files), which are then executed on available nodes. The configuration of available nodes is reported in yet another log file.

Therefore, our simulation framework has to cope with several independent sources of system configuration and must process them in synchronized manner (all entries in traces files come with timestamp or period range). There are four main sources of configuration state changes in our workload simulation model:

- Dynamic resource usage of processes – the resources utilization levels are not constant through the life on an application and sometimes vary greatly from their specified requirements.
- New jobs are scheduled and current jobs complete their processing or are cancelled – this is the core operation in any scheduled system. When a given job is scheduled, user specifies required resources and constraints on node attributes.
- Changes in jobs resource requirements and/or constraints – during execution, tasks might have their resource requirements and constraints altered. This may result in a node being no longer suitable for certain type of task.
- Changes in nodes configurations – during a cluster system lifecycle, nodes might be taken offline for maintenance or upgraded, with new nodes being added or old nodes removed. This scenario is rarely visible in smaller data centers (i.e. MetaCentrum infrastructure is relatively infrequently updated [8]), however it is more common to find frequent alteration to configuration within larger systems such as Google Cluster.

To be able to handle this highly concurrent environment, all workload state updates (new tasks, updated constraints, new nodes, removed nodes, etc.) are done via immutable events as shown in Fig. 1.

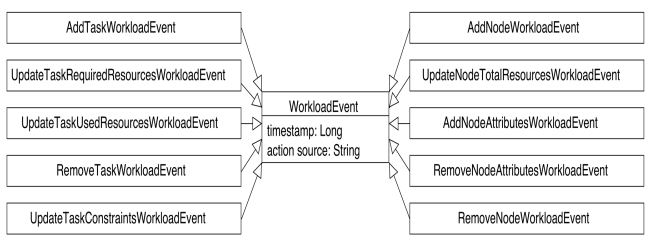

Figure 1. Workload events

Every event is marked with a timestamp to sort event batches from several parsers in correct execution order. Such a setup enables the simulation system to maintain consistency of state even under heavy load. Detailed descriptions of all workload events are presented below:

- AddTaskWorkloadEvent – This event is generated for each new task. Tasks are always generated with initial resource requirements and constraints. However, resource requirements and constraints might change during execution of task and thus render node unsuitable for processing this task.
- UpdateTaskRequiredResourcesWorkloadEvent – In the majority of cases, requested resources values do not change after initial value. However, in several instances tasks get their required resources updated. This event is initially generated together with AddTaskWorkloadEvent, however it will also be generated upon update of task required resources.
- UpdateTaskUsedResourcesWorkloadEvent - Upon execution, tasks dynamically allocate various amounts of memory; utilize storage space in different levels, etc. This event is generated to keep track of currently allocated resources.
- UpdateTaskConstraintsWorkloadEvent - Task constraints are a set of logical operators set on node attributes and their values, which enable or disable execution of that task on certain node. Similarly to the task's required resources, task constraints are dynamic values and can be updated at any time.
- RemoveTaskWorkloadEvent - This event is generated when the task finishes its execution or is killed by system or user. When examining Google workload traces, we have noticed that significant parts of the tasks were killed by the native system.
- AddNodeWorkloadEvent - This event is generated when a new node is added to the cluster. The majority of this kind of events is generated upon the start of the simulation.
- UpdateNodeTotalResourcesWorkloadEvent - During the simulation lifecycle, certain nodes are taken offline and their resources updated (i.e. new memory banks are added). As with AddNodeWorkloadEvent, the majority of this kind of event is happens at the start of the simulation.
- AddNodeAttributesWorkloadEvent - Similarly to node resources, certain attributes can be also updated. GCD project doesn't specify the meaning of attributes (values and names are obfuscated), however it suggests features like: existence of external IP address, specific version of the Linux kernel, etc. This event is generated if the node attributes are updated or new attributes are added.
- RemoveNodeAttributesWorkloadEvent - As in the event above, node attributes can be removed (i.e. meaning that the node no longer has an external IP address).
- RemoveNodeWorkloadEvent - Upon simulation certain node were taken down (for maintenance of completely removed from cluster). This event removes node from available nodes pool.

GCDs' traces keep record of all tasks updates and action. A task might have only two states with a number of transformations between those states: pending (task is awaiting allocation to a node) and running (task is running on a node). Once allocated, the running task cannot go back to pending state. If task is evicted or lost a clone task is created added to the queue. Fig. 2 below presents the lifecycle of a task and shows how those updates are directly mapped to workload events in the simulator.

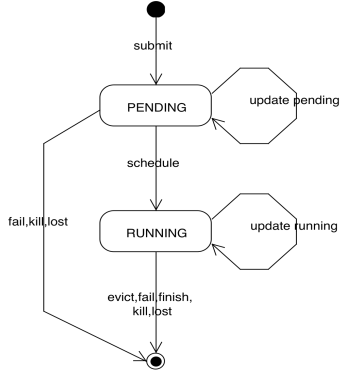

Figure 2. Task lifecycle

Transformation between states generates relevant WorkloadEvents as summarized in Table I below:

TABLE I.  TASKS TO WORKLOAD EVENTS MAPPING

| Action | Description | Workload Event |
|---|---|---|
| SUBMIT | Task has been submitted to cluster scheduler | AddTaskWorkloadEvent |
| SCHEDULE | Task has been scheduled by proprietary Google cluster scheduler | (no event generated) |
| EVICT | Task has been evicted (and killed) from node. | RemoveTaskWorkloadEvent |
| FAIL | Task failed (i.e. execution crashed) | RemoveTaskWorkloadEvent |
| FINISH | Task finished normally | RemoveTaskWorkloadEvent |
| KILL | Task has been killed (by user or system) | RemoveTaskWorkloadEvent |
| LOST | Task was terminated, but there is no record indicating that | RemoveTaskWorkloadEvent |
| UPDATE PENDING | Task resource levels or constraints were updated | UpdateTaskRequiredResourcesWorkloadEvent |
| UPDATE RUNNING | Task resource levels or constraints were updated during execution | UpdateTaskRequiredResourcesWorkloadEvent |

The above task actions impact the workload state:
- SUBMIT action creates task in queue and sends it to Workload Manager. This is where initial task resource requirements are registered in AddTaskWorkloadEvent;
- SCHEDULE actions are results of actions of internal Google scheduler, therefore simulator is ignoring them;
- EVICT, FAIL, FINISH, KILL and LOST actions mark the end of a task and are translated to RemoveTaskWorkloadEvent;
- UPDATE PENDING and UPDATE RUNNING actions mark changes in task priority, required resources and task constraints. Thos are often results of users changing requirements of already submitted jobs. The UpdateTaskRequiredResourcesWorkloadEvent is generated;
- Changes in task constraints are independently managed via UpdateTaskConstraintsWorkloadEvent.

The key components of AGOCS framework are shown in Fig. 3. The simulator uses five independent workers implemented as Actors from Akka framework, which read and parse workload traces data files and generate Workload Events.

Each Events Parser holds a buffer of events (30 min ahead of simulation time or hard limit of 1000000 events) to avoid synchronous methods' calls. When a worker remains idle and the system usage is low, it will fill the events buffer.

Proper timing of task state transformations is very critical to simulation. Therefore we have designed central module with responsibility to ensure all events are read and processed in the same time window. Every five seconds the WorkloadGenerator collects events from the events parsers and updates the system state in a shared system object ContextData. This object is repeatedly read by various system elements; therefore it has been designed to be able to support highly concurrent scenarios.

All workload states are stored in a set of thread-safe lock-free implementations of a hash array mapped trie [3] – Scala's TrieMap structure [17]. The workload simulator is highly concurrent – all blocking operations are wrapped within Scala Futures, which move computations to another thread, while the main part of the code is executed in parallel. Such design approach allowed us to fully utilize all available CPU cores on the testing machine.

Due to the size of the archive (191GB), we could not fit it into memory and have decided to continuously keep reading and parsing trace files at runtime, while keeping a set of events in fast-access memory.

The main purpose of these buffers is to minimize blocking operation while reading and preparing the next set of events. Each events parser keeps a buffer of events in memory (30 minutes of events ahead; no more than 1 million events) and releases them to the Workload Manager on request (every 5 seconds). If event parser does not have requested set events, it will block the request (synchronous call) until enough events are loaded from data files. While idle, each events parser will passively keep reading data files and parsing them into buffered events.

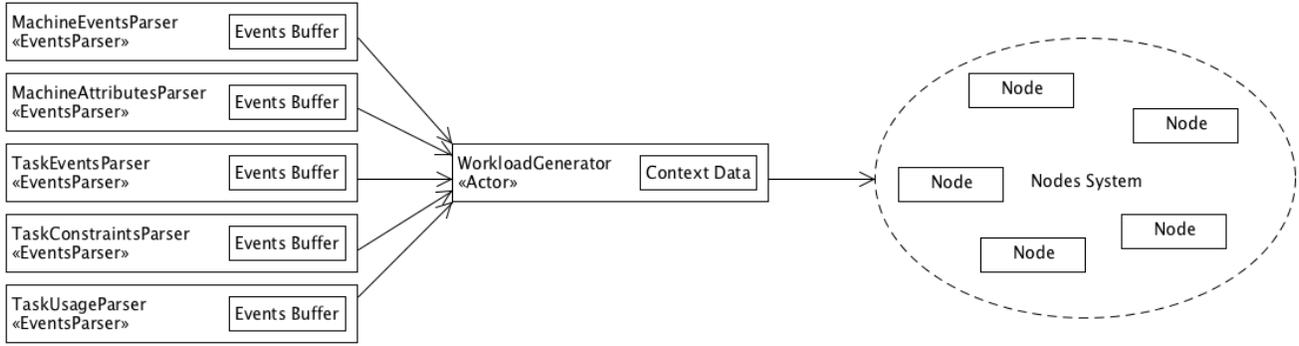

Figure 3. Workload events generation

## IV. USE CASE

AGOCS was created as a part of a research project Multi-Agent System Balancer (MASB) [21, 22], whose aim is to design and implement an intelligent scheduler for a Cloud system. Cloud systems are very complex with dynamic entities containing many overlapping dependencies; therefore providing a reliable Cloud load simulation with high quality workload traces is of upmost importance.

As already described in Section III, the presented framework generates and accurately times workload events and feeds them to designed service's destination (push model). The typical architecture for testing consists of a single stand-alone AGOCS server (not necessary on a separate machine) and a number of simultaneously running instances of schedulers as shown in Fig. 4.

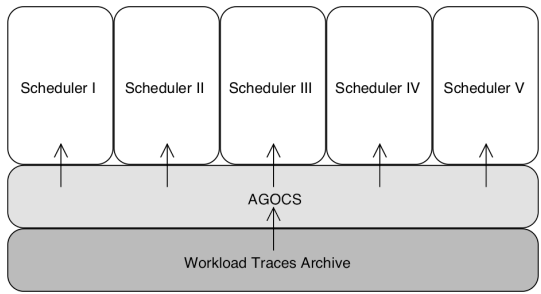

Figure 4. AGOCS use case

An interesting feature of AGOCS is that it can be paused at any time, allowing users to take a snapshot of current tasks distributions and state of scheduled jobs (loading current state snapshot and restarting simulation is not implemented yet). This approach enables researchers to conveniently and directly compare various scheduling algorithms at any time while they are running, without being limited to output results and statistics.

While most of simulation framework functionalities are enabled in configuration files or triggered by command line, AGOCS also offers graphical simulation monitor module (Fig. 5). Currently, simulation monitor needs to be run in the same environment as framework server, but a fully detachable and stand-alone monitor application will be created in the future.

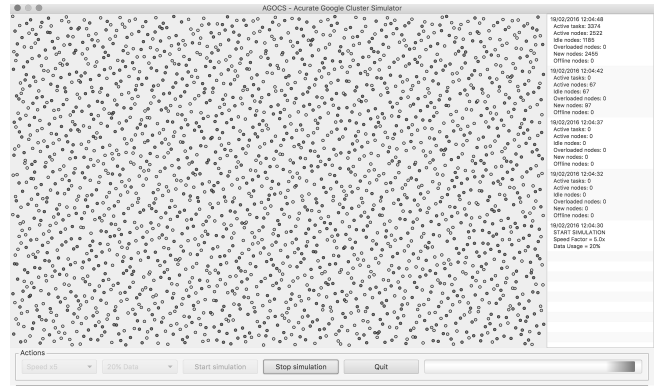

Figure 5. AGOCS simulation monitor

AGOCS was designed for common desktop machines; however it does require loading and processing a huge amount of workload traces data (uncompressed 191GB). Therefore, running a simulation framework server is disk I/O-intensive and may interfere with OS's swap memory operations. Nevertheless, we were able to work around this issue by attaching an external disk drive to our testing machine. Currently examined distributed scheduling algorithms are fairly CPU-intensive (some are also memory-intensive), therefore a more powerful machine is required to run a large number of simulations in parallel. Nevertheless, this lightweight design allows us to comfortably run month-long simulation on our testing machine (Apple MacBook Pro 2.7GHz dual-core Intel Core i5 8GB RAM) in ca.9 hours with 75x speed factor, which is equal to processing ca.21.22GB of data per hour. The majority of data (ca.89%) of data (170.54GB) comes from resource usage log files. After initial loading and buffering of data (ca.30 seconds), system runs with consistent ca.10-15% CPU usage.

Currently, in the MASB project testing simulation consists of up to 5 concurrently deployed meta-heuristic schedulers selected from the following: Greedy, Tabu Search, Simulated Annealing and 4 variations of the Genetic Algorithm with and without seeding as described in [22]. More schedulers are to be implemented in the future. All schedulers are processing the same workload events from a single AGOCS server. However, the speed factor is set only to five times faster processing due to our current hardware processing power limitations.

## V. ALTERNATIVE DESIGNS

During our experiments so far, the design of the simulation framework significantly evolved based on the project requirements and experiences. While Scala and other high-level languages offer a variety of out-of-the-box routines and functions (especially related to concurrency), it is sometimes better to implement certain functionalities to maintain better control of data flow. For the AGOCS simulator we designed in fine-details the process of generating and maintaining workload state, while leaving to Akka the task of running our processes in parallel. Akka Actors are designed for highly concurrent event-driven applications and can be deployed in distributed environment. Therefore, AGOCS can be deployed on multiple machines.

Nevertheless, there are several optimizations and strategies that could be applied to build an even more efficient workload simulator. In the sub-sections below we list alternative design approaches.

### A. Pre-processing of data files

The current simulator implementation reads directly from the GCD work traces files. However, we were using only a fraction of all available data. It is especially visible when reading and parsing task resource usage (task_usage) files, where we disregard majority of fields (i.e. mapped/unmapped page cache memory usage, mean and maximum disk I/O time, cycles per instruction, etc.). The simulator flow could be refactored into two stages:

- First, where we read and process all available workload traces data and store them into a list of events. Result list of events could be persisted to file or other persistence media.
- Second, where workload generator reads and replays records of previously stored events.

Such an approach would help us avoid processing massive amounts of original data and also significantly reduce the overhead from parsing logic. However, the trade-off would be higher complexity of the code and less flexibility during experiment. Pre-processed data could be stored either in a set of files or a database. Storing data in database would provide an additional benefit of a mature query APIs such as SQL or NoSQL, which could be used to directly examine workload data by external applications.

### B. Streaming events generation

Java 8 and also Scala since it is JVM-based provide various features and optimizations for streaming operations [23]. Most implemented transformations such as parsing files, filtering bogus events, and sorting by timestamp could be natively converted into parallel operations, concurrently executed on all available CPU cores. Such a stream would be split into separate pipelines and each event will be created and examined as a separate process. The OS would execute each pipeline in parallel on multiple CPU cores and then their results would be collected into a single queue.

It is difficult to estimate the performance gains from this approach; however it could significantly reduce complexity of existing code base. The trade-off would be less control over generation of events since proper timing and staging is difficult to achieve when framework controls the utilization of pipelines. Nevertheless, it is a very viable approach and might greatly reduce the management logic overhead for timing and caching of workload events in simulations.

The flow of proposed stream operations is shown below (Fig. 6). Workload Events stream would start with reading and streaming lines from all archive files in parallel (*flatMap* operation). Generating a single Workload Event might require combining several lines from several files (i.e. UpdateTaskConstraintsWorkloadEvent) and this logic is encapsulated in the Event Parsers (*map* operation). Each Event Parser generates a uniform Workload Event object, which then is accepted or denied by the Events Filter (*filter* operation). Filtering events is critical to avoid bogus data. Finally, a collection of those objects is ordered by timestamp (*sortBy* operation) and fed to the Workload Generator (*collect* operation), which then distributes them to destination services via configured protocol.

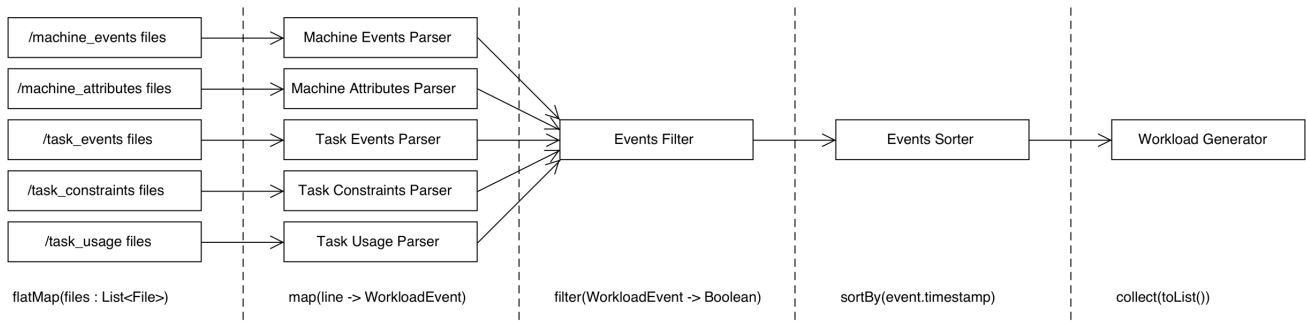

Figure 6.  Stream-based simulator

## VI. RELATED WORK

Currently there exist a number of Cloud simulation frameworks such as CloudAnalyst [25], GreenCloud [13], Network CloudSim [11, 14] or EMUSIM [7, 14]. Those frameworks were designed to cover a wide range of Cloud systems simulations, while AGOCS was designed with a focused goal of simulating Google Cloud cell environment with consideration of a very fine-grained and detailed aspect simulation such as tasks resource utilization, task constraints, jobs queue simulation, jobs and tasks priority class, node's local scheduler simulation, detailed statistic as memory cache hit ratio, etc. A brief comparison between those frameworks is shown in Table II below. It should also be noted that while the AGOCS framework has currently implemented parsers only for Google Data Cluster project, parser classes could be easily extended to support other workload traces format. We estimate that experienced developer should be able to finish this task within a week.

At the same time, AGOCS was designed with usability in mind. As a very lightweight framework it is capable of being run on typical desktop machine available in any laboratory.

During our experiment we have found that we can comfortably run month-long simulation on our testing in ca.9 hours with 100x speedup factor, which is equal to processing ca.21.22 GB of data per hour. Table II below presents a brief comparison between the AGOCS framework and a few selected Open Source simulators [14].

While in this section we present only brief comparison to a range of existing Cloud simulations, in the next section we focus on evaluating the performance of AGOCS against its closest alternative – the CloudSim package.

TABLE II. CLOUD SIMULATORS COMPARISON

| Framework | **AGOCS** | **CloudAnalyst** | **GreenCloud** | **Network CloudSim** | **EMUSIM** |
|---|---|---|---|---|---|
| Platform | Scala/Akka | CloudSim | NS2 | CloudSim | AEF |
| Language | Scala | Java | C++/OTCL | Java | Java |
| Simulator Type | Event Based | Event Based | Packet Level | Packet Level | Event Based |
| Supported workload traces | Google Cluster Data (CSV) | Custom (ASCII/XML) | Loadable configuration settings (TCL) | Custom (ASCII/XML) | Custom (ASCII/XML) |
| Networking | Limited | Limited | Full | Full | Limited |
| Resource constraints | Yes | Yes | Yes | Yes | Yes |
| Supported and reported resource types | - CPU Cores (Requested/Used)<br>- Canonical Memory (Used)<br>- Assigned Memory (Requested/Used)<br>- Page Cache Memory (Used)<br>- Disk I/O Time (Used)<br>- Local and Remote Disk Space (Requested/Used)<br>- Cycles Per Instruction (Used)<br>- Memory Access Per Instruction (Used)<br>- Local Scheduler (Priority Class)<br>- Jobs and Tasks Priority | - CPU Cores (Requested)<br>- Bandwidth (Requested)<br>- Memory (Requested)<br>- Millions of Instructions Per Second | - Server Load factor (Requested/Used)<br>- Bandwidth (Requested/Used)<br>- Memory (Requested/Used)<br>- Energy Used (split by servers, switches, etc.)<br>- Service Timeout | - CPU Cores (Requested)<br>- Bandwidth (Requested)<br>- RAM size (Requested)<br>- Millions of Instructions Per Second | - CPU Cores (Requested)<br>- Bandwidth (Requested)<br>- RAM size (Requested)<br>- Millions of Instructions Per Second |
| Attribute constraints | Yes | Limited | No | Limited | Limited |
| Build-in scenarios | Google Cluster (cell A), 12K nodes | Generator and examples | Examples | Generator and examples | Several pre-defined scenarios |

## VII. PERFORMANCE EVALUATION

The closest alternative Cloud simulator to AGOCS is CloudSim package [11], created in Melbourne 'Clouds' Lab and currently available as Open Source project in GitHub repository [1]. For the purpose of our experiment we have retrieved and compiled the latest version available (version 3.1-SNAPSHOT, last commit dated 8th Feb 2016).

Both tools are created with JVM-based technologies and therefore the testing environment is identical – Java SE Runtime Environment 1.8.0_60 run on OS X 10.11.3 (El Capitan) on MacBook Pro11.1 with 2.4GHz dual-core Intel Core i5 and 8GB RAM.

Generally speaking, CloudSim offers greater flexibility when setting up an environment. Nodes and tasks (referred to as 'cloudlets' in CloudSim package) are set up in Java classes, which are then compiled to separate jar package and run together with the main jar file. This approach is advantageous as compiled classes can be further automatically optimized by JVM even during execution (HotSpot technology). On the other hand, AGOCS is not configured statistically, but continuously reads workload traces files and updates its state. That ensures simulation is very scalable, however, those parsing operations are quite expensive and have impact on performance.

To have comparable input data sizes, in our tests we have configured both frameworks to run the same number of tasks and nodes. On average during simulation GDC schedules ca.140k tasks on ca.12.5K nodes, thus we generally preserved this ratio (i.e. 11 tasks per node, e.g. 5500 task were submitted to 500 nodes) in performance evaluation. We have also configured CloudSim to assign a single VM to a single host machine. AGOCS has been run with the highest possible speed factor that our testing machine was capable of running. Fig. 7 below presents the simulation time results.

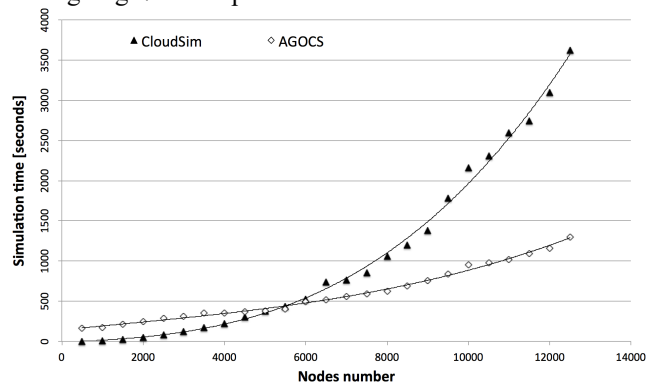

Figure 7. AGOCS performance evaluation and comparison with CloudSim

CloudSim performs better in smaller sets of data, however the computation time increases significantly for more complex sets. AGOCS's computation time increase is less rapid, though it requires initial time to pre-load data to its buffers. AGOCS was designed with multi-threading in mind and can take advantage of all available CPU cores. Its main bottleneck is workload traces' reading speed. CloudSim is completely memory-driven; however it is implemented as a single-threaded application and utilizes only one CPU core. CloudSim's code is prone to over-use of Java's ArrayList class, while HashSet would work much faster – significant amount of CloudSim's simulation time is spent in Java's ArrayList.removeAll method. On the whole, AGOCS simulates more layers of complexity and it is more accurate when describing secondary machine parameters:

- AGOCS supports adding and removing nodes during simulation. In smaller Cloud systems [8] this may not be an issue as machine configuration is very static, however in bigger environment nodes are frequently modified and/or exchanged [14].
- Tasks being executed have actual values for both requested and actually used resources; this is very important factor as task resources utilization level is not constant and usually fluctuates depending on activity tasks currently performed. Therefore, to obtain a realistic and current view on node machine utilization we need to consider actually used resources, not requested ones. It should also be noted that CloudSim framework also supports several resource utilization models defined on task: full, stochastic and pre-defined (based on PlanetLab datacenter's traces [11]).
- Tasks and nodes are representation of real machines and tasks run on Google Cluster. While CloudSim can generate random parameters based on statistical analysis, this approach will widen error margins and uncommon machine configurations might be missed.
- AGOCS simulation provides not only values for defined resources, but also a number of secondary parameters, such as: disk I/O time, cycles per instruction, memory access per instruction, etc. This improves realism of simulation and might serve as a input for processes.
- AGOCS's tasks have sets of constraints and nodes have sets of attributes. While node has enough resources to run certain tasks, it might be missing some features required to successfully complete the task completely (e.g. availability of external IP address).

Therefore, as demonstrated above, AGOCS provides more complex and accurate simulation than CloudSim package at the expense of performance and flexibility. However, AGOCS has limitations and constraints, that researchers should be aware of:

- While providing relatively detailed description of physical layer and requested resources as well as rich set of secondary parameters, AGOCS does not provide values for bandwidth utilization. Unfortunately, GDC workload traces do not provide values for network transfer and this could be a critical missing feature in some research projects.
- The CloudSim package is easily extendable by a third party and several other tools have been already built upon this framework [11, 14, 25], often adding new features and new resource types. AGOCS is based on already existing workload traces and until Google decides to repeat this experiment (and potentially extend the set of monitored parameters), new modules are highly unlikely.
- The main risk of using AGOCS for research is that generated workload does not change between interactions. Simulation is replayed the same way every time, i.e. timings of tasks, changes of nodes, etc. are always identical. This may lead to developing over-specialized or over-trained algorithms that work on a single particular set of data only.

However, the length of provided traces (one-month) is more than enough to evaluate researched product in wide variety of situations. Nevertheless, the researcher has to be aware of the above limitations in an attempt to achieve an accurate simulation.

AGOCS and CloudSim share many similarities and features, but are representing quite different approaches to the same research problem. Our experience is related to designing scalable load balancing strategies for Cloud systems. CloudSim provides informative top view of a Cloud system and is strong in testing high-level algorithms and strategies, while AGOCS is very suitable to fine-tune those algorithms and running simulations that are as close as possible to real Cloud systems. Load balancing strategies need to consider very fine-grained details and effects, often originating from the physical layer of a tested system. Due to implemented complexity, AGOCS is very appropriate for this class of research.

## VIII. CONCLUSIONS

There are certain aspects of the design of the workload simulator that can be noted:

- Simulating Cloud workloads on a complex network is not simplistic. Significant number of parameters and dependencies require a well-designed domain model. Especially implementation of constraints logic is critical and time-consuming.
- The use of real-world traces improves the accuracy and guarantees the generation of realistic results by the simulator. Generating random (but still statistically valid) workload data is suitable for experiments with high-level algorithms. However, some cases of problems (e.g. load balancing or task scheduling), researcher must consider very fine-grained parameters and effects, often originating in physical layer of Cloud system.
- The system should be able to cope with data anomalies and data corruption. The available traces are of high quality, but anomalies exist in provided data. Common data errors include: corrupted state of task (i.e. task is marked as running when job has already finished) or corrupted usage logs (i.e. reporting task resource usage for non-existing task).
- In many cases it is not feasible to fully load all simulation data to memory. Therefore the simulator framework has be able to process external files (or storage) and update its state on the fly.
- Due to the complexity of data, it is difficult to properly test the created simulator. The design should allow for simulator 'testability' in mind. Every object and state should have appropriate unit tests during implementation.

AGOCS continues to be actively developed and new features are continuously added. Recent updates have enabled the system to distribute over a number of machines and speed up the simulation. Future work on the framework will involve a visualization module of Cloud nodes workload and we are considering implementation of parsers for alternative workload traces formats. Additionally, the framework's roadmap includes creation and loading of task distribution and workload state snapshots, which will help during debugging and analysis for further development and improvement of scheduling algorithms.